\documentstyle[prb,aps,preprint]{revtex}
\begin{document}
\draft
\def\overlay#1#2{\setbox0=\hbox{#1}\setbox1=\hbox to
\wd0{\hss #2\hss}#1%
\hskip -2\wd0\copy1}
\title{
Screening effect of interlayer coupling on \\
electronic transport properties
of layered high-$T_c$ oxides
}
\author{Ryoko Sugano, Toshiyuki Onogi, and \\Yoshimasa Murayama}
\address{
Advanced Research Laboratory, Hitachi, Ltd.,
 Hatoyama, Saitama 350-03, Japan
}

\maketitle

\begin{abstract}
We have investigated the effect of thermal fluctuation
on the interlayer coupling
of three-dimensional Josephson-junction arrays with
anisotropic interactions
as a model of layered high-$T_c$ oxides,
focusing on non-Ohmic current-voltage characteristics.
Langevin dynamic simulations were performed
for various Josephson-coupling anisotropies
and for various temperatures
in both zero and finite magnetic fields.
We find that,
in the highly anisotropic region,
the interlayer coupling,
which causes the non-vanishing critical current ($I_{c}$),
is suppressed by thermal fluctuations, and
tends to push up
the Kosterlitz-Thouless transition temperature ($T_{KT}$)
of the two-dimensional (2D) regime.
In this highly anisotropic region,
$I_c$ begins to decrease drastically near $T_{KT}$.
This suggests that interlayer coupling is screened by
thermally excited 2D small vortex-antivortex pairs as well as
the intralayer logarithmic part of vortex-antivortex interactions.
Moreover, weak magnetic fields parallel to the $c$-axis
affect only the logarithmic-interaction part
between the 2D pancake vortices.

\end{abstract}



\narrowtext

\section{introduction}

The coherence length ($\xi_{ab}$)-scale thermal fluctuation
in high-$T_c$ oxides
has been extensively discussed in association with
the thermal excitation of
two-dimensional (2D) vortex-antivortex pairs.
In the strictly 2D case, it is well known that
the logarithmic interaction between these vortex-antivortex pairs
shows power-law behavior of
current-voltage ($I$-$V$) characteristics, and
leads to the Kosterlitz-Thouless (KT) transition.\cite{kosterlitz73}
Recently, the KT transition has been observed even in
a one-unit-cell(12\AA)-thick YBCO thin film.\cite{matsuda92}
This clearly indicates that
superconductivity in high-$T_c$ oxides
is intrinsically two-dimensional.
Under weak magnetic fields,
power-law behavior and the KT transition have also been observed in
both $I$-$V$ characteristics and magnetoresistance.\cite{martin89}
These observations were also confirmed by
numerical simulations of 2D Josephson-junction (JJ) arrays.
\cite{sugano92}

On the other hand,
high-$T_c$ superconductors in bulk are
three-dimensional (3D) weakly coupled
stacks of superconducting CuO$_2$ planes.
The existence of
a weak but non-zero interlayer Josephson-coupling
between CuO$_2$ layers has been observed.\cite{farrell89}
This coupling could alter the logarithmic interaction
between 2D pancake vortices,\cite{minn90}
thus blunting transition,
shifting the Kosterlitz-Thouless transition temperature $T_{KT}$,
\cite{matsuda93,hikami80} and
modifying $I$-$V$ characteristics.\cite{jen91,sugano93}
Moreover,
the correlation of 2D vortices in adjacent layers
is expected to decrease
near the transition temperature $T_{cr}$, and
it has been observed that
layers are decoupled when $T > T_{cr}$.
\cite{minn91,schilling93,eltsev94,safer92}
It is not yet known whether layers are
completely decoupled just below $T_{cr}$.

In this paper, we systematically investigate
the screening effect of the interlayer coupling on
the electric transport properties
in both zero and weak magnetic fields ($H \!\parallel c$),
and discuss the results based on the KT mechanism.

\section{model and formalism}

We start with a 3D
anisotropic Josephson-junction (JJ) array
regarded as a simple lattice version of
a type-II superconductor derived
from a Lawrence-Doniach type model.\cite{doniach71}
The Hamiltonian is written as
\begin{equation}
{\cal H}\!=\!-J_{\perp c}\!
\sum_{<i,j>_{\perp c}}\cos( \Delta \theta_{ij} - A_{ij})
-J_{\parallel c}\!\sum_{<k,l>_{\parallel c}}\!
\cos( \Delta \theta_{kl} - A_{kl}),
 \label{hami}
\end{equation}
where $\Delta\theta_{ij}$ denotes
the superconducting phase difference
between nodes $i$ and $j$.
$J_{\perp c}$ and $J_{\parallel c}$ are
intralayer (inplane) and
interlayer Josephson-coupling energies related to
the magnetic penetration depth
$\lambda_{\parallel c,\perp c}$, respectively,
such that $J_{\parallel c,\perp c}$=
${\Phi_{0}}^2a/16\pi^3{\lambda^2}_{\parallel c,\perp c}$,
and also related to
the bare critical current of each Josephson-junction through
$I_{0\parallel c, \perp c}$=($2e/\hbar$)$J_{\parallel c, \perp c}$.
Here $a$ and $\Phi_0$ are a lattice spacing and
flux quantum, respectively.
We assume that $a$ is comparable to
the inplane coherence length $\xi_{ab}$.
Bare anisotropy ratio $\gamma_0 = \gamma(T=0)$ at T=0 is defined as
\begin{equation}
\gamma_0 = \gamma(T=0) = {J_{\parallel c}/ J_{\perp c}}=
(\lambda_{\perp c} / \lambda_{\parallel c})^2
\label{anis}
\end{equation}
At finite temperature,
$\gamma_0$ can be replaced by
effective anisotropy ratio $\gamma(T)$.
Magnetic fields are introduced as frustration
$2\pi f=A_{ij}+A_{jk}+A_{kl}+A_{li}$ around each plaquette
with $A_{ij}=(2e/\hbar c){\int_i^j}{\bf A}\!\!\cdot\!\!d{\bf l}$
and $f=Ha^2/\Phi_0$.
The external DC current $I$ is injected by adding
$-(\hbar/2e)I$$\sum_{<i,j>_{\parallel I}}\!\Delta\theta_{ij}$
to Eq.\ (\ref{hami}).
We directly calculated the $I$-$V$ characteristics of
the 8 $\times$ 8 $\times$ 3 JJ array
for various reduced temperatures $k_BT/J_{\perp c}$, and
for various weak magnetic fields,
via Langevin simulation technique,
as previously reported in Ref[10].

\section{vortex dynamics}

Thermal excitation of 2D vortex-antivortex pairs plays
an important role in 2D superconductivity ($\gamma_0=0$).
Such a 2D vortex pair separated by
distance $r$ within the same layer
interacts with a logarithmic potential
$U_{2D}=E_1 \ln (r/\xi_{ab})$,
and leads to the simple power-law $I$-$V$ characteristics:
\begin{equation}
R=V/I \propto I^{\alpha-1}.
\label{iv2d}
\end{equation}
Moreover, in this 2D regime, the KT transition may be caused by
spontaneous unbinding of vortex pairs due to thermal fluctuation.
At the KT transition temperature $T_{KT}$,
the critical singularity can be observed as
a discontinuous jump (called universal jump)
from 2 to 0
in the temperature dependence of the power-law exponent $\alpha-1$.
This jump separates
the non-Ohmic resistive state ($\alpha-1 > 2$) for $T<T_{KT}$
from the Ohmic state ($\alpha-1=0$) for $T>T_{KT}$.\cite{kadin83}
In the presence of
a weak interlayer Josephson-coupling ($\gamma_0 \neq 0$),
it should be noted that the interlayer coupling modifies
the logarithmic interaction between vortices
within the same layer.\cite{minn90}
The modified vortex-antivortex interaction energy
$U_{mod}(r)$ is distinguished
by the characteristic size of vortex pairs $r_{\gamma}(T)$,
which is defined by
effective anisotropy ratio $\gamma$ based on Eq.\ (\ref{anis}):
\begin{equation}
r_{\gamma}(T) = \xi_{ab}/\sqrt{\gamma (T)}.
\label{r0}
\end{equation}
Here the strong anisotropy regime is defined by $r_{\gamma} \gg 2a$,
i.e., $\sqrt{\gamma} \ll 0.5$, where
the thermally excited 2D vortex fluctuations are dominant rather than
the 3D vortex loop fluctuations.\cite{sugano93}
In this regime ($\sqrt{\gamma} \ll 0.5$),
while on the one hand
the size $r$ of vortex pairs smaller than $r_{\gamma}$
adds the $r$-square interaction
to the logarithmic interaction between vortices, \cite{hikami80}
the larger size ($r \gg r_{\gamma}$) adds
the $r$-linear interaction.\cite{minn91}
\begin{eqnarray}
U_{mod}(r) &=& E_c + E_1(T) \ln (r/\xi_{ab})
               + (1/4) \gamma(T) E_1(T) (r/\xi_{ab})^2
          \qquad   ( r \ll r_{\gamma})  \label{rsmall} \\
     &=& E_c + E_1(T) \ln (r/\xi_{ab}) +
            (\pi/\sqrt{2}) \sqrt{\gamma(T)}E_1(T)(r/\xi_{ab}-1)
             \qquad  ( r \gg r_{\gamma})
\label{rlarge}
\end{eqnarray}
Here $E_c$ is the energy to create a 2D vortex in one layer, and
$E_1(0)=2\pi J_{\perp c}$.
Compared with the second logarithmic term and
the third $r$-square term in Eq.\ (\ref{rsmall}),
the logarithmic term is considerably larger than the $r$-square term,
and thus 2D behavior is displayed for the smaller-$r$ regime
($r \ll r_{\gamma}$).
For the larger-$r$ regime ($r \gg r_{\gamma}$), on the other hand,
the $r$-linear term is of the same order as the logarithmic term
in Eq.\ (\ref{rlarge}),
and we should expect the interlayer Josephson coupling effect
to appear clearly in the electric properties.

Considered with the $r$-linear form of the Lorentz force,
this added $r$-linear interaction,
based on large-size vortex pair excitations,
modifies the $I$-$V$ characteristics
with non-vanishing critical current $I_c$
in the larger-$r$ regime, \cite{jen91}
and this is supported by
numerical simulations of 3D anisotropic JJ arrays.
\cite{sugano93}
\begin{equation}
  R \propto (I-I_c)^{\alpha-1}.
     \label{modiv}
\end{equation}
Here,
\begin{equation}
I_c \sim \sqrt{\gamma(T)} c E_1(T) /\Phi_0 \xi_{ab}
\label{icg}
\end{equation}
and
\begin{equation}
\alpha-1 \sim E_1(T)/k_{\em B} T
\label{ag}
\end{equation}
are required.
In strongly anisotropic BSCCO
with $\gamma \sim 4 \times 10^{-4}$,
$r_{\gamma} \sim 1000$ \AA~\ is significantly large,
and almost pure 2D behavior is expected to appear.
On the other hand, in less anisotropic YBCO
with $\gamma \sim 4 \times 10^{-2}$,
$r_{\gamma} \sim 100$ \AA, and
this much smaller value of $r_{\gamma}$ is considered to
greatly enhance the $r$-linear term,
even for the 10-unit-cell(120 \AA)-thick film.

Focusing on this larger-$r$ regime,
we below studied
the effect of the interlayer coupling
on the electric properties
in zero and finite magnetic fields for various finite temperatures.
In particular, we will discuss
the power-law $I$-$V$ characteristics with
a non-vanishing critical current $I_c$ and
the power-law exponent $\alpha-1$
in Eqs.\ (\ref{modiv})-(\ref{ag}),
which eventually result in $\sqrt{\gamma(T)}$ and
$E_1(T)$ renormalized with
the screening effect of the interlayer coupling
by 2D vortex-antivortex fluctuations in the weak coupling regime.

\section{result and discussion}

The inset of Fig.~\ \ref{iv} shows log-log plots of
$I$-$R$ ($R=V/I$) curves for
various reduced temperatures $k_{B}T/J_{\perp c}$
at $H=0$ and $\gamma_0=0.04$.
On the low temperature side ($k_BT/J_{\perp c} \leq 0.7$),
$I$-$R$ curves display a negative curvature,
and seem to obey the modified power-law with
a non-vanishing critical current $I_c$
in Eq.\ (\ref{modiv}),
whereas, in the higher temperature regime
($k_BT/J_{\perp c} \geq 0.7$),
they tend to accord with
the simple power-law relationship in Eq.\ (\ref{iv2d}).
Also, in this high temperature regime
the power-law exponent $\alpha -1$ seems to be
smaller than 2.
We can also see this feature more clearly
by replotting these curves together with Eq.\ (\ref{modiv})
as depicted in Fig.~\ \ref{iv}.
Here the dotted lines are fitted for
each $I$-$R$ curve using on $\chi^2$ fit based on Eq.\ (\ref{modiv}).
The $I$-$R$ curves seem classified into two groups.
One group, for $k_BT/J_{\perp c} \leq 0.7$,
has non-Ohmic power-law behavior
with a non-vanishing critical current $I_c$ and
with an exponent which $\alpha-1$ larger than 2.
The other is for $k_BT/J_{\perp c} \geq 0.7$ and
displays nearly Ohmic power-law behavior for $\alpha-1 < 2$.

In the presence of external current,
the modified interaction energy between two vortices
from Eq. \ (\ref{rlarge}) is written by
\begin{eqnarray}
U_{mod}(r) &=& E_c+E_1 \ln (r/\xi_{ab}) + E_2(r/\xi_{ab} -1)
         - (\Phi_0/c)Ir  \label{nonumber} \\
     &\sim& E_c+E_1 \ln (r/\xi_{ab}) -
            (\Phi_0/c) (I -I_c) r,
\label{lorent}
\end{eqnarray}
where
$E_2 = \sqrt{\gamma}E_1$.
For the strictly 2D case at $\gamma(T)=0$,
the third term in Eq.~\ (\ref{nonumber}) vanishes because $E_2=0$,
and Eq.~\ (\ref{nonumber}) results in the pure 2D interaction of
the logarithmic form.\cite{kadin83}

Due to the third term in Eq.\ (\ref{lorent}) for $\gamma_0 \neq 0$,
current-induced unbinding of 2D vortex pairs
are suppressed by
\begin{equation}
I_c = E_2(c/\xi_{ab}\Phi_0) = \sqrt{\gamma} E_1 (c/\Phi_0 \xi_{ab}).
\label{ice2}
\end{equation}
The maximum value of $U_{max}$ and
the maximum size $r_{max}$ ($U_{max}=U_{mod}(r_{max})$),
that give the thermal activation rate $\Gamma$ from bound pairs
to free vortices by $\Gamma \propto e^{-U_{max}/k_{B}T}$,
become larger than in the 2D case.
Therefore the resistivity is suppressed in the presence of
the interlayer coupling.

Figure~\ \ref{Ict} (a) shows the temperature dependence of $I_c$
for various values of bare anisotropy ratio $\gamma_0$ at $H$=0.
$I_c$ decreases with increase in temperature.
In particular, $I_c$ vanishes in the high temperature region
for the extremely weak coupling ($0 < \gamma_0 < 0.1$).
The effect of the interlayer coupling is
suppressed by thermal fluctuations, and vanishes to zero.
When we compare the $I$-$R$ curves and
the temperature dependence curve of $I_c$ for $\gamma_0=0.04$,
we can see that Ohmic characteristics with $\alpha-1 < 2$
appear in the region where $I_c$ vanishes.
There are two different ranges distinguished by
temperature $T_{Ic}$ with $I_c(T>T_{Ic})=0$.
Moreover $I_c$ is extremely suppressed just below $T_{Ic}$.
These regions are believed to be associated with the KT transition,
as we will discuss later.

In the temperature dependence of $\alpha-1$
as shown in Fig.~\ \ref{Ict} (b),
we can see a kink-like anomaly
in the curve for the 2D case ($\gamma_0=0$)
around $\alpha-1 = 2$.
A discussion of this anomaly was based on
the KT vortex-antivortex unbinding transition
corresponding to a sign for the universal jump.\cite{kadin83}
For the modest-size (8$\times$8$\times$3) lattice, unfortunately,
numerical evidence for the KT singularity is inconclusive,
since the universal jump is expected to broaden largely,
according to $\Delta T \propto 1/\ln (L)$ with linear size $L$.
{}From the numerical data,
we extract the KT transition temperature ${T_{KT}}^{2D}$,
tentatively determined by the exponent $\alpha({T_{KT}}^{2D})-1=2$
for the 2D case at $\gamma_0 =0$.
Even for the weak coupling case ($0 \neq \gamma_0 <$ 0.1)
we also see the analogous kink-like anomalies around $\alpha-1=2$.
This suggests that the KT transition would survive
in the presence of such a weak interlayer coupling.
The temperature at the anomaly ${T_{KT}}^{3D}(\gamma_0 \neq 0)$
seems to increase monotonically with increasing $\gamma_0$.
Moreover, each ${T_{KT}}^{3D}$ corresponds to
each vanishing temperature $T_{Ic}$
in the temperature dependence of $I_c$ (Fig.~\ \ref{Ict} (a)),
that is, the KT anomaly appears at the time $I_c$ vanishes
(${T_{KT}}^{3D}(\gamma_0)=T_{Ic}(\gamma_0)$ for $\gamma_0 \neq 0$).
Since $I_c(T) \propto E_2(T)$,
both the intralayer logarithmic interaction part $E_1(T)$ and
the interlayer $r$-linear part $E_2(T)$ of Eq.~\ (\ref{nonumber})
must be screened by 2D vortex-antivortex fluctuations.
This screening effect of vanishing to zero of $E_2(T)$
above ${T_{KT}}^{3D}$
is consistent with Monte Carlo results by Minnhagen et. al.
\cite{minn91}
It is suggested that
the system is anisotropic 3D below ${T_{KT}}^{3D}$
and is 2D just above ${T_{KT}}^{3D}$
($E_2(T) \neq 0$ for $T<{T_{KT}}^{3D}$,
and $E_2(T) = 0$ for $T>{T_{KT}}^{3D}$).

Note that there are plateau regions
in the temperature dependence of $\alpha-1$
just below ${T_{KT}}^{3D}$
for the case with the non-zero interlayer coupling.
The size of the plateau becomes larger
with the increase of the interlayer coupling.
The plateau begins
at the KT anomaly temperature ${T_{KT}}^{2D}$ without
the interlayer coupling around $\alpha-1 = 2$.
Just above that temperature ${T_{KT}}^{2D}$,
$I_c$ begins to decrease drastically with increase in temperature.
This rapid decrease in $I_c$ can be regarded as
the drastic suppression of effective anisotropy ratio $\gamma(T)$.
This is confirmed by seeing
the temperature dependence of $\sqrt{\gamma(T)}$
given by the ratio of
the exponent $\alpha-1$ to the critical current $I_c$:
\begin{equation}
\sqrt{\gamma(T)} \sim {E_2(T) \over E_1(T) }=
{\Phi_0 \xi_{ab} \over {c k_B}}{I_c \over {(\alpha-1)T}}.
\label{xgt}
\end{equation}
Here we use Eq.~\ (\ref{ag}) and Eq.~\ (\ref{ice2}).
Figure~\ \ref{xg} shows the temperature dependence of
our extracted value of $\sqrt{\gamma(T)}$ in Eq.~\ (\ref{xgt})
at $H=0$ for various bare anisotropy ratios $\gamma_0$.
We can see temperature independent behavior
in the low temperature phase ($T < {T_{KT}}^{2D}$),
i.e., $\gamma(T)$ has a constant value below ${T_{KT}}^{2D}$.
This implies that below ${T_{KT}}^{2D}$
both $E_2(T)$ and $E_1(T)$ are renormalized with
the same form, and
effective anisotropy ratio $\gamma(T)$ is not screened
($\gamma(T)=\gamma_0$ for $T<{T_{KT}}^{2D}$).
$\gamma(T)$ begins to decrease toward zero above ${T_{KT}}^{2D}$.
The rapid decrease of $\gamma(T)$ in the high temperature region
agrees with Pierson's Renormalization Group (RG) study
\cite{pierson94} including the screening
due to small intralayer 2D vortex pairs.
Small 2D vortex pairs in each plane are also relevant
to the weakening of the $r$-linear part of
the interaction in the high temperature phase,
where the logarithmic part of the interaction is fully renormalized.
This also indicates that correlation along the $c$-axis is reduced.

The qualitative change around ${T_{KT}}^{3D}$ is interpreted as
a sort of dimensional crossover
from anisotropic 3D vortex fluctuations caused by
the Josephson coupling between layers to
pure 2D vortex fluctuations of the decoupled regime.
The decoupled regime above ${T_{KT}}^{3D}$ exhibits Ohmic-behavior,
while non-Ohmic power-law $I$-$R$ curves of
$R \propto (I-I_c)^{\alpha-1}$ appear
in the weak coupling regime below ${T_{KT}}^{3D}$.
Due to the modest-size (8$\times$8$\times$3) lattice,
our calculated power-law exponent $\alpha-1$ is not exactly 0,
but $\alpha-1 <2$, in the decoupled regime.
Nevertheless here it is interesting that
smaller vortex pairs reduced both interlayer $r$-linear and
intralayer logarithmic interactions between 2D vortex pairs;
full renormalization of
the logarithmic part of interaction seems to be
a trigger for reduction of the $r$-linear part of the interaction.

We can also see the power-law behavior of $I$-$R$ curves,
fitted with Eq.\ (\ref{modiv}), even in the weak magnetic field
at $k_{B}T/J_{\perp c} =0.6$ and $Ha^2/\Phi_0 = 0.0104$,
in Fig.~\ \ref{power} (a).
Log-plus $r$-linear interaction between 2D vortex pairs survives
even in the magnetic field.
Moreover, there is other power-law behavior in magnetoresistance,
as depicted in Fig.~\ \ref{power} (b),
at $k_{B}T/J_{\perp c} =0.6$ and $I/I_{0 \perp c}=0.4$.
\begin{equation}
R \propto H^{\beta}
\label{hr}
\end{equation}
The coexistence of two power-law behaviors
(Eq.~\ (\ref{modiv}) and Eq.~\ (\ref{hr})) leads mathematically to
the logarithmic dependence of the power-law exponent $\alpha-1$
on the magnetic fields as in Fig.~\ \ref{exp}:
$\alpha-1 \propto \ln 1/H$. \cite{martin89,sugano92}
The external field suppresses the exponent $\alpha-1$ through
the logarithmic function of the magnetic field $Ha^2/\Phi_0$.
The focusing points appeared in both magnetoresistance and
the magnetic field dependence of the exponent $\alpha-1$.
These focusing points have
the almost same value of $Ha^2/\Phi_0$ of the order of 0.015.
With our assumption that $a \sim \xi_{ab}$,
$\xi_{ab} = 12$ \AA~\ in YBCO gives $H \sim 20$ T, and
$\xi_{ab} = 38$ \AA~\ in BSCCO gives $H \sim 2$ T.
This suggests that
the power-law behavior in YBCO is kept up
under ten times higher magnetic fields than that in BSCCO.
This is consistent with experimental results.\cite{martin89}
In the weak-field regime $Ha^2/\Phi_0 < 0.015$
(Fig.~\ \ref{power} (b) and Fig.~\ \ref{exp}),
the picture in terms of field-induced unbinding of 2D vortex pairs
is relevant even in the strongly anisotropic system
($\sqrt{\gamma_0} < 0.5 $).
The inset of Fig.~\ \ref{exp} shows
effective anisotropy ratio $\sqrt{\gamma(T)}$
at $k_{B}T/J_{\perp c}$ =0.6 as a function of $Ha^2/\Phi_0$.
As the magnetic field increases,
square-root effective anisotropy ratio
$\sqrt{\gamma(T,H)} \sim E_2(T,H)/E_1(T,H)$
seems to increase slightly, at $k_BT/J_{\perp c}$=0.6.
{}From the relation $\alpha-1 = E_1(T,H)/k_BT$ in Eq.~\ (\ref{ag}),
it is suggested that the external field suppresses
only the logarithmic part of interaction $E_1(T,H)$,
and maintains or enhances the $r$-linear part due to
the interlayer coupling
$E_2(T,H) \simeq \sqrt{\gamma(T,H)}E_1(T,H)$.
We can also see this tendency in Fig~\ \ref{gdep}, where
$I_c \propto E_2$ certainly obeys
the $\sqrt{\gamma(T,H)}$-dependence, and
$E_1(T,H)$ (the slope of the curve) screened by
the magnetic field is expected to give a larger value of
$\sqrt{\gamma(T,H)} \sim E_2(T,H)/E_1(T,H)$
than that in the absence of the field.
As a result, effective anisotropy ratio $\gamma(T)$
seems to be enhanced in the $\gamma$-constant phase
($T < {T_{KT}}^{3D}(\gamma_0,H)$),
that is, the weak magnetic field would apparently
strengthen the interlayer coupling.
{}From this larger value of $\gamma(T)$,
we expect to find that
the plateau in the temperature dependence of $\alpha-1$,
where $I_c$ drastically decreases,
is larger under the weak magnetic field
than in the absence of any field.
Because of this clearer plateau,
the universal jump could survive in the magnetic field,
as observed experimentally.\cite{martin89}

In summary,
we have performed Langevin simulations of
the 3D anisotropic JJ arrays
for a model of layered high-$T_c$ oxides.
We found that the effect of vortex fluctuation due to
small inplane vortex pairs effectively weakens
the interlayer coupling and wipes it out, that is,
effective anisotropy ratio $\gamma(T)$ is screened by
thermally excited 2D vortex-antivortex pairs in each layer.
Just above ${T_{KT}}^{3D}$, layers are decoupled, and
the system becomes 2D.
This is consistent with
Monte Carlo study by Minnhagen and Olsson \cite{minn91} and
with RG study by Pierson \cite{pierson94}.
Signs of the transition seem to appear
around the temperature $T_{I_c}$ where the critical current $I_c$ vanishes.
Moreover, we can see that ${T_{KT}}^{3D}(\gamma_0)$ tends to
shift slightly up in the high temperature direction
for small $\gamma_0 \neq 0$.
When the interlayer coupling becomes
small for $0 < \gamma_0 < 0.1$,
$I_c$ vanishes
in the high temperature region, and
this vanishing of $I_c$ may play
a role in triggering the KT transitions.
Since $\gamma(T)$ drastically decreases
with increase in temperature
just below ${T_{KT}}^{3D}$,
such a weakening effect of the interlayer coupling must
mainly be due to screening by the high density of 2D vortex pairs.
Thus, at around ${T_{KT}}^{3D} = T_{I_c}$,
there appears a sort of dimensional crossover
from the 3D anisotropic system to the pure 2D system.
Magnetic fields affect only the logarithmic interaction
between 2D vortex-antivortex pairs.
In our simulation, we have not introduced
the effect of the coherence length $\xi_{c}$.
Near $T_c$, significant growth of $\xi_{c}$
beyond the spacing between CuO$_2$ layers
is expected to give
less anisotropic or isotropic 3D behavior far above ${T_{KT}}^{3D}$
for $\sqrt{1/(1-T/T_c)} \gg 1$.
The KT transition observed in YBCO appears to be
a sort of dimensional crossover from the anisotropic 3D system to
the strict 2D system,
while the one observed in BSCCO is
the typical KT vortex-antivortex unbinding transition
in the pure 2D system.
At that point, YBCO in bulk can be distinguished from BSCCO.

\acknowledgments

We are grateful to Prof. S. Hikami, Prof. Matsuda,
Prof. H. Kawamura, Dr. T. Ohta,
and Dr. Hirao for fruitful discussions.
Numerical calculations were carried out on a HITAC S-820
supercomputer at the Advanced Research Laboratory of Hitachi, Ltd.

\newpage

\begin{figure}
\caption{Resistance $R=V/I$ as a function of $I-I_c$
at $\gamma_0=0.04$ for various values of
reduced temperature $k_{\rm B}T/J_{\perp c}=0.6$
in the absence of a magnetic field.
Inset: resistance $R$ as a function of external current $I$.}
\label{iv}
\end{figure}

\begin{figure}
\caption{Temperature-dependence for various values of
bare anisotropy ratio $\gamma_0$:
  (a) Critical current $I_c$ as
a function of temperature $k_{B}T/J_{\perp c}$,
  (b) Power-law exponent $\alpha-1$ as
a function of temperature $k_{B}T/J_{\perp c}$.}
\label{Ict}
\end{figure}

\begin{figure}
\caption{Temperature dependence of $E_2(T)/E_1(T)$
for various values of bare anisotropy ratio $\gamma_0$.
($E_2(T)/E_1(T)$ is
proportional to the square-root of $\gamma(T)$)}
 \label{xg}
\end{figure}

\begin{figure}
\caption{Power-law type behavior
for various values of bare anisotropy ratio $\gamma_0$
at $k_BT/J_{\perp c}=0.6$:
 (a) Power-law magnetoresistance at current $I/I_0=0.4$,
 (b) Power-law type $I$-$R$ curves
at magnetic field $H a^2/ \Phi_0 =0.0104$.}
 \label{power}
\end{figure}

\begin{figure}
\caption{
Logarithmic behavior of the power-law exponent $\alpha -1$
at $k_BT/J_{\perp c}=0.6$;the exponent $\alpha-1$
as a function of magnetic field $H a^2/ \Phi_0$.
Inset: $ E_2 / E_1 $
as a function of $H a^2 / \Phi_0$.}
 \label{exp}
\end{figure}

\begin{figure}
\caption{Anisotropy dependence at $k_BT/J_{\perp c}=0.6$
for various magnetic fields;
critical current $I_c$ as a function of
$\protect\sqrt{\gamma(T)}$
with effective anisotropy $\gamma(T)$. }
 \label{gdep}
\end{figure}

\hspace{.5cm}


\newpage

\begin{references}

\bibitem{kosterlitz73}J. M. Kosterlittz and D. J. Thouless,
J.\ Phys.\ {\bf C 6}, 1181 (1973);
B. I. Halperin and D. R. Nelson,
J.\ Low\ Temp.\ Phys.\ {\bf B36}, 599 (1979).

\bibitem{matsuda92}Y. Matsuda, et. al.,
Phys.\ Rev.\ Lett.\ {\bf 69}, 3288, (1992);
T. Terashima et al.,
Phys.\ Rev.\ Lett.\  {\bf 67}, 1362 (1991).

\bibitem{martin89}S. Martin et al,
Phys.\ Rev.\ Lett.\  {\bf 62}, 677 (1989);
M. Ban, T. Ichiguchi, and T. Onogi,
Phys.\ Rev.\ {\bf B40}, 4419(1989);
T. Onogi, T. Ichiguchi, and T. Aida,
Solid\ State\ Commun.\ {\bf 69}, 991(1989).

\bibitem{sugano92}R. Sugano, T. Onogi, and Y. Murayama,
Phys.\ Rev.\ {\bf B45}, 10789 (1992);
R. Sugano, T. Onogi, and Y. Murayama,
Physica\ {\bf C 185-189}, 1643 (1991).

\bibitem{farrell89}D. E. Farrell et al.,
Phys.\ Rev.\ Lett.\  {\bf 63}, 782 (1989);
R. Kleiner et al.,
Phys.\ Rev.\ Lett.\  {\bf 68}, 2394 (1992).

\bibitem{minn90}V. Cataudella and P. Minnhagen,
Physica\ {\bf C 166}, 442 (1990);
S. N. Artemenko and A. N. Krugluv,Physica {\bf C173}, 125 (1991),
M. V. Feigelman, et al.,Physica {\bf C167}, 177 (1990).

\bibitem{matsuda93}Y. Matsuda, et. al.,
Phys.\ Rev.\ {\bf B48}, 10489, (1993);
T. Ota, et al., Phys.\ Rev.\ {\bf B50}, 3363, (1994).

\bibitem{hikami80} S. Hikami and T, Tuneto,
Prog.\ Theor.\ Phys.\  {\bf 63}, 387 (1980).

\bibitem{jen91} H. J. Jensen and P. Minnhagen,
Phys.\ Rev.\ Lett.\ {\bf 66}, 1630 (1991).

\bibitem{sugano93} R. Sugano, T. Onogi, and Y. Murayama,
Phys.\ Rev.\ B{\bf 48}, 13784 (1993).

\bibitem{minn91} P. Minnhagen and P. Olsson,
Phys.\ Rev.\ {\bf B44}, 4503 (1991);
P. Minnhagen and P. Olsson,
Phys.\ Rev.\ Lett.\  {\bf 67}, 2596 (1991).

\bibitem{schilling93} A. Schilling, et al.,
Phys.\ Rev.\ Lett.\  {\bf 71}, 1899 (1993).

\bibitem{eltsev94} Yu. Eltsev, W. Holm, and O. Rapp,
Phys.\ Rev.\ {\bf B49}, 12333 (1994).

\bibitem{safer92} H. Safer, et al.,
Phys.\ Rev.\ {\bf B46}, 14238 (1992).

\bibitem{doniach71}W. E. Lawrence and S. Doniach,
in {\em Proceedings of LT12, Kyoto, 1970},
edited by E. Kanda (Keigaku, Tokyo, 1971), p. 361.

\bibitem{kadin83} A. M. Kadin, K. Epstein, and A. M. Goldman,
Phys.\ Rev.\ {\bf B27}, 6691 (1983).

\bibitem{pierson94} S. W. Pierson,
Phys.\ Rev.\ Lett.\  {\bf 73}, 2596 (1994).

\end{references}
\end{document}